# Equilibrium route to colloidal gellation: mixtures of hard sphere-like colloids

Ph. Germain and S. Amokrane.

Laboratoire de Physique des Liquides et Milieux Complexes, Faculté des Sciences et Technologie, Université Paris Est (Créteil), 61 Avenue du Général de Gaulle, 94010 Créteil Cedex, France.

**Abstract :**

The binodals and the non-ergodicity lines of a binary mixture of hard sphere-like particles with large size ratio are computed for studying the interplay between dynamic arrest and phase separation in depletion-driven colloidal mixtures. Contrarily to the case of hard core plus short range effective attraction, physical gellation without competition with the fluid-phase separation can occur in such mixtures. This behavior due to the oscillations in the depletion potential should concern all «simple» mixtures with non-ideal depletant, justifying further studies of their dynamic properties.

The question of gellation in soft condensed matter systems has been the subject of a lively debate in the past few years [1-8]. On the one hand, following the extension of the conceptual framework developed for ordinary glasses [9,10], a broad picture of the arrest in soft matter ones has emerged from the recent literature [11]. In comparison with the "repulsive" glasses in which the arrest is driven mostly by repulsions ("caging" mechanism), soft matter ones exhibit also a transition to different glassy states driven by short range effective attractions ("bonding" mechanism). At least in the effective fluid approach, such effective attractions, defined in a broad way, exist in a variety of colloidal suspensions, as for example, globular proteins, colloidal silica or colloid-ideal polymer mixtures, etc (see for example respectively refs. [12,13], [14] and [15]). Despite their quite different origins and shapes, these attractions could lead to attractive glasses – or gels at low density- provided that their range is short enough.

On the other hand, the interplay between such arrested states and the equilibrium phase transitions in colloidal systems is not completely established. More precisely, the possibility for colloids with short range attraction to form equilibrium physical gels – that is reversible gels in the equilibrium fluid phase – has been questioned, since this attraction plays also a role in the phase separation. In order to analyze these questions, different theoretical studies have been performed. Most of them used generic model potentials that retained only what seemed the most significant feature of the effective interaction: the short range, irrespective of its precise shape. For such simplified models (square well, Yukawa potential, etc. ), however, increasing the attraction strength also favors the fluid-fluid transition. The question is then to know if gellation should occur outside the fluid-fluid coexistence domain, setting aside a possible crystallization.

Early studies based on the mode coupling theory (MCT) [9] suggested that gellation is indeed possible without an intervening phase separation, when the attraction range is sufficiently short [8,11,16]. Similar observations were also made on less asymmetric mixtures [17,18] (these however fall in a different category, for which a multicomponent approach is necessary). Nevertheless, subsequent numerical simulations showed that the glass line intercepts in fact the fluid-fluid binodal for such short range potentials (see [1] for a review). From the ensuing scaling with the attraction range of the dynamics and of the static transition lines, a new paradigm was developed according to which physical gellation – associated with short range attraction – would only be observed through an arrested fluid–fluid phase separation [1-3]. It should apply to all the systems for which this picture of the interactions – essentially a hard core repulsion and a very short range attraction – holds as globular proteins, colloidal silica or colloid-ideal polymer mixtures mentioned above (the latter systems have been the subject of numerous studies as the depletion interaction can then be tuned easily). Very recent experiments seem to corroborate this paradigm [13,19,20] though some previous studies [6,7,21] are compatible with the opposite view. This motivated the search of more complex interaction in order to favor equilibrium gellation (long-range electrostatic repulsion [22], "patches" [23], etc.).

The purpose of this letter is to revisit this view focused on "short range attractive colloids " and the resulting correlation between gellation and phase separation. We will first show that while it is natural to consider the effective interaction in these systems as essentially a hard core plus a short range attraction, this is in fact insufficient for discussing the interplay between dynamic arrest and phase instability: without considering complicated models, we will illustrate on the example of an asymmetric binary mixture of hard spheres that the characteristics of the effective potential (range, depth, repulsive barriers, etc.) can affect differently the non-ergodicity and phase transition lines. While preserving gellation, this may go up to suppressing the fluid-fluid coexistence and hence the very question of their competition. This considerably widens the scenarios for the gel transition and its interplay with the equilibrium phase transitions in colloidal systems.

Accordingly, wee will show that pseudo-binary mixtures of hard colloids might constitute good candidates for fulfilling the conditions for equilibrium gellation. Indeed, while these mixtures are very simple, the effective interaction potential is there naturally complex enough: after the well at contact, it shows pronounced oscillations at the scale of the small hard spheres. These features are well known but it has become customary in the literature to neglect them and to reduce the depletion effect to the short range attraction, exemplified by the Asakura-Oosawa (AO) potential for the colloid-*ideal* polymer mixture. We will show that these oscillations typical of real "depletion" mixtures can actually have a strong influence on the extent of the non-ergodicity domain: one fundamental effect of the repulsive barriers is to stabilize the physical "bonds"; furthermore, this goes for some size ratios with a simultaneous removal of the fluid-fluid transition because of a particular combination of the strength and range of the depletion well. From this analysis we will predict the size ratios for which equilibrium gellation should be observed.

We now briefly describe the theoretical method used here (see also [24]). The binodals and the non-ergodicity transition lines have been computed for a model binary mixture, in the effective one-component fluid (EOCF) representation. The thermodynamic variables are the big particle packing fraction $\eta_b$, the small particles density in the reservoir $\rho_s^*$ and the temperature $T$. The total effective potential between the colloids is the sum of the direct interaction potentials and the indirect one, computed at infinite dilution from the RHNC (reference hypernetted chain) integral equation. The phase diagram of the EOCF is computed in the $(\eta_b, \rho_s^*)$ plane, using the hybrid RHNC/variational perturbation treatment for the binodals and the MCT for the non-ergodicity transition line. For the static properties, the accuracy of this method has been established for the size ratios investigated here (see e. g. [25-27]).

The one-component MCT has been widely for studying the repulsive and attractive arrested states in hard particle systems (for reviews see [11] and [1]). Its use for the study of gellation in mixtures raises two different questions. The first one concerns the reliability of the MCT transition line in the vicinity of the fluid-fluid phase transition, due to the critical density fluctuations (see for a general discussion [8] and [28] for molecular dynamics simulations). This should not however constitute a real problem here, since we precisely consider situations in which there is no fluid condensation. In the same situations, the question of cluster aggregation (see e. g. [7], [1]) that occurs at very low density should also not arise. The second question concerns the reduction of the mixture to an effective one component fluid for studying gellation. One qualitative argument is that for the size ratios and the densities considered here ($\eta_s^* \leq 0.4$) the fluid of small hard spheres should remain ergodic in the free volume. More quantitatively, there are partial indications in favor of this from recent studies of star polymer mixtures [29] and simulations [30]. Now, the very question of the use of an effective potential to study the dynamic arrest at low density is also substantiated by general arguments such as the adiabaticity criterion [31] but this question remains open. Finally, due to the absence of fluid condensation in all the situations we discuss here, a possible involvement of the small spheres – above some packing fraction – in the arrest should make that gellation would take place at even lower packing fraction of the big ones that in the one-component representation. This would just reinforce our conclusions on gellation in these systems.

We first present the phase diagram of the hard sphere (HS) mixture with diameter ratio $q = \sigma_2/\sigma_1 = 5$ (figure 1). A comparison with the Asakura-Oosawa (AO) model is also presented (inset) for the fluid-fluid transition and the non-ergodicity lines. We observe for the HS mixture a quite distinct pattern, in comparison with the AO or other generic short-range potentials: first, in the HS mixture, the fluid-fluid binodal is absent (none is found at size ratios $q \leq 8$ [26, 32]) and second, the non-ergodicity line is shifted to lower packing fractions. For $\rho_s^* = 0.8$, we find $\eta_g = 0.32$ and 0.19 for $q = 5$ and 8 respectively. Since the fluid-fluid instability is absent, one should thus observe gellation in the equilibrium homogenous fluid, provided that crystallisation is prevented by a small amount of polydispersity [33]. This is in sharp contrast with the behaviour found with the AO model, for which the metastable fluid-fluid binodal exists, and the non-ergodicity transition is confined to the dense fluid region (the behaviour observed with the present MCT for $\rho_s^* > 0.6$ should be taken with care, as it occurs just below the fluid-fluid transition [7, 8]). These features are typical of usual models of attractive potentials with moderate range [8, 16]: while the fluid-fluid binodal is metastable with respect to the fluid-solid one due to the short range, this range is not short enough to induce low density gellation. In the HS mixture, on the contrary, gellation is observed, while the fluid condensation disappears.

This specific behaviour of the mixture of hard colloids may now be correlated with the characteristics of the hard-sphere depletion potential. The latter originate from the repulsion between the small hard particles, ignored in AO model. Its interplay with that between the small and the big ones leads to a more complex behaviour of the HS depletion potential, $\phi_{HS}^{eff}$. In place of a single well with range $1/q$, $\phi_{HS}^{eff}$ is oscillatory with repulsive barriers and wells varying in a complex way with $q$ and $\rho_s^*$. Concerning the gellation line, one expects some influence of the repulsive barriers of $\phi_{HS}^{eff}$ (figure 2). The most important one is located right after the depletion well. Its magnitude $\Delta\varepsilon^{rep}$ becomes comparable to the depletion well depth, $\Delta\varepsilon^{att}$, when $\rho_s^*$ increases. The consequences on gellation are shown in the inset of figure 1 which compares the non-ergodicity transition lines for the full depletion potential $\phi_{HS}^{eff}$ and for a truncated version $\phi^*$ (without barriers): $\phi^*(x \leq \delta) = \phi_{HS}^{eff}(x)$ and $\phi^*(x > \delta) = 0$, with x=(r-$\sigma_b$)/$\sigma_b$ the reduced distance and $\delta^*$ the reduced width of the attractive well. For $\phi^*$, the non-ergodic state is confined to the dense region, as for the AO potential. The repulsive barrier favors thus arrest at lower density. A simple interpretation of this observation is that the barrier stabilizes the "bonds" by making more difficult for the particles to escape from the depletion well. For example, for $\rho_s^*$=0.8, the energy $\Delta\varepsilon = \Delta\varepsilon^{att} + \Delta\varepsilon^{rep}$ associated to such a bond is $\Delta\varepsilon \approx 5.4\,k_BT$, instead of $\Delta\varepsilon \approx 3\,k_BT$ for $\phi^*$ (figure 2). This interpretation is substantiated by the behavior of the localization lengths $r_{loc}$ of the two models (computed here from the simplest gaussian approximation $f_q \approx \exp(-q^2 r_{loc}^2/6)$ of the non-ergodicity factor

[6]): along the non-ergodicity transition line, $r_{loc}$ is indeed systematically smaller with $\phi_{HS}^{eff}$ than with $\phi^*$.

Now, is the repulsive barrier the unique ingredient to stabilize equilibrium gellation ? In addition to a short range attraction, the role of similar barriers has indeed already been pointed out for this purpose. They are for example artificially introduced in numerical simulations performed for generic potentials for stabilizing homogenous gellation *against* fluid condensation [34]. Besides the fact that the barriers are here real, an important difference is that, in our case, they are irrelevant for the question of the fluid condensation which is absent both with $\phi^*$ and with $\phi_{HS}^{eff}$. This is due to the reduction with $\rho_s^*$ of the width of the attraction well (inset, figure 2): for. $\rho_s^* = 0.8$ for e. g, this width is about four times smaller than its low density limit, $\delta^{(0)} \approx \frac{1}{q}$, the AO density independent value. This is why the fluid condensation is absent for $\phi_{HS}^{eff}$ and not for $\phi_{AO}$. Homogenous gellation in the hard-sphere mixture results thus from a subtle mechanism: on the one hand, the repulsive barrier favors arrest at lower density by stabilizing the bonds, and on the other hand, the fluid condensation is suppressed by the simultaneous reduction with the small particles density of the width of the attraction well.

Both in order to test the robustness of these conclusions and anticipate the behavior of real suspensions, we have considered mixtures of hard spheres with a very short range tail in the interaction potential between unlike ones: $\frac{V_{sb}(r)}{k_B T} = \varepsilon^* \frac{\sigma_{sb}}{r} \exp(-\frac{r - \sigma_{sb}}{\xi_{sb}})$. We took values typical of "residual" interactions in hard sphere-like colloids (say via the surface layers as in sterically stabilized ones [35], see [36] for details): $\varepsilon^* = \pm 1.5$ and $\xi_{sb} = \frac{\sigma_s}{100}$. With $q = 5$, this corresponds for example to $\sigma_s$=0.2µm, $\sigma_b$=1µm and $\xi_{sb}$ = 2 nm [35]. Such *a priori* "small" interactions can have in fact important consequences on the binodals at high size asymmetry

(see [36] for q = 10). We show in figure 3 the situation for q = 5. The F – F transition remains absent, as with pure hard spheres. The gel line is moderately shifted towards lower (greater) values of $\eta_b$ according to the sign of $V_{sb}$. This is a natural consequence of the enhancement ($V_{sb}, > 0$) or the reduction by solvation ($V_{sb} < 0$) of the depletion mechanism. This does not modify qualitatively the picture relative to pure hard spheres, contrarily to the case $q = 10$. On this basis, it seems reasonable to predict that an equilibrium gel can form in mixtures of hard sphere-like colloids with moderate asymmetry ($q \sim 5$), irrespective of the details of the residual interactions (sterically stabilized or screened charged ones with very small screening range).

In summary, we have studied the interplay between the arrested states at low density and the phase separations for mixtures of hard colloids. It is shown that the oscillations with separation between the particles of the effective interaction potential can be responsible for quite specific behaviors. In particular, the repulsive barriers have been shown to provide a stabilizing mechanism of the physical bonds involved in gellation. As the fluid-fluid phase transition is not observed for some size ratios, gelation is predicted to occur without competition with the fluid condensation. Thus, the behavior of simple asymmetric mixtures can depart from that expected from the "hard core short-range attraction" picture. This observation should stimulate reconsideration at the experimental level of these already known systems (see [37]). They indeed have been much less considered in the literature than colloid-polymer mixture, perhaps because of the greater convenience of using polymers as the depletant. At the theoretical level, additional simulations should be useful to assess the validity of the methods used to study gellation, and in particular the quantitative predictions made here from the one-component mode coupling theory. If confirmed, it should have practical applications, besides the additional insight it provides on the mechanisms of arrest in soft condensed matter systems.

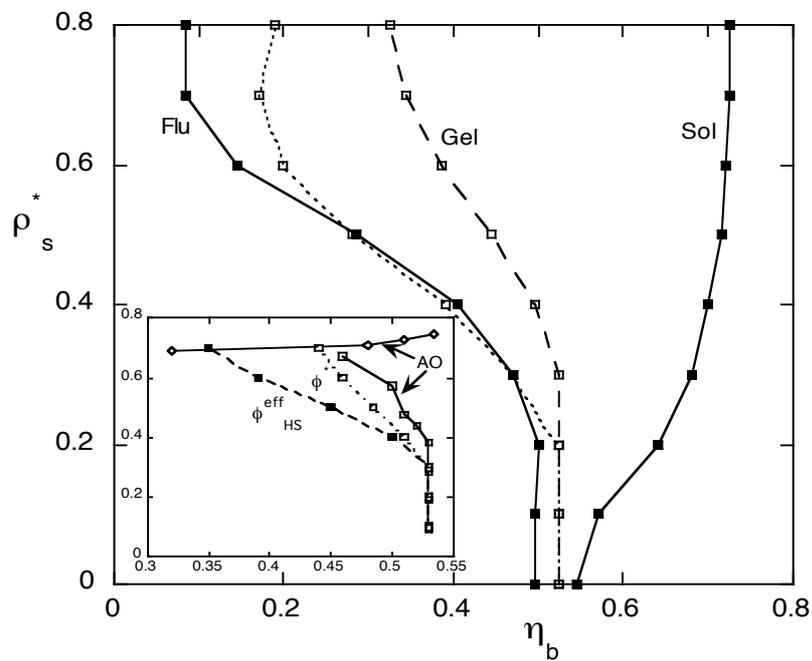

**Figure 1 :**

Phase diagram of the HS mixture for q=5, in the effective one-component fluid representation. Solid lines : binodals ; Dashes : non ergodicity line (dots: q = 8). Inset : fluid-fluid binodal and non ergodicity lines for the Asakura-Oosawa (solid line), the HS depletion potentials (long dashes), and for the truncated potential $\phi^*$ (short dashes). The fluid-fluid binodal (diamonds) is present only for the AO model.

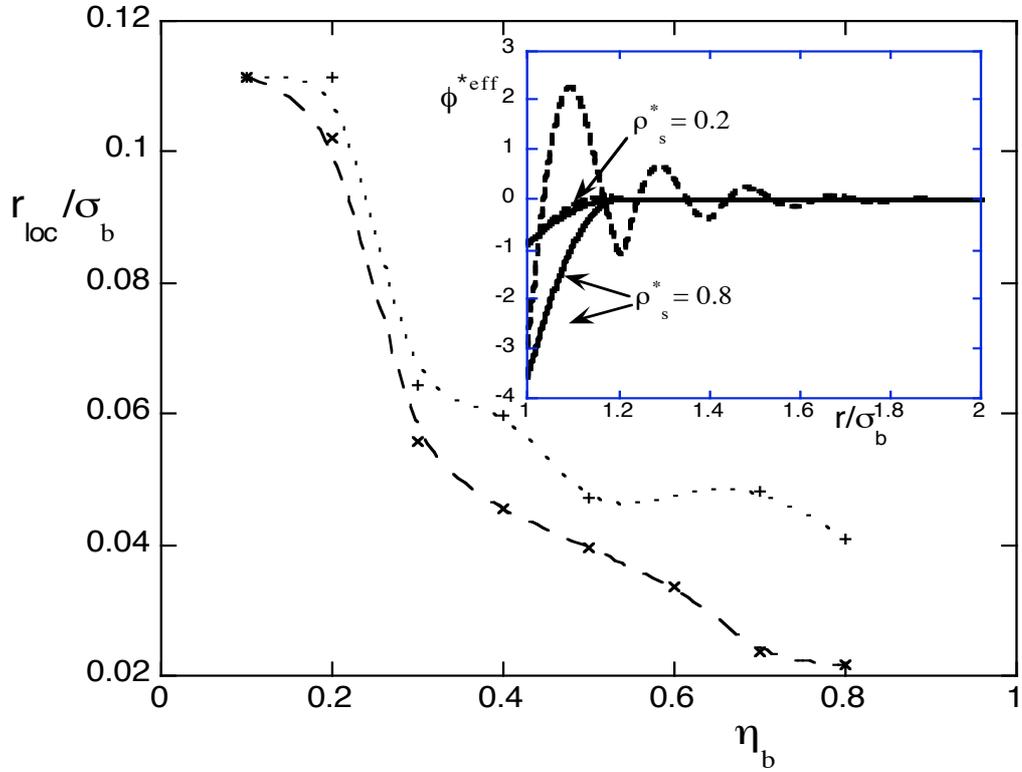

**Figure 2 :**

Localization length of the big particles in the gel state along the non-ergodicity transition line, for $q = 5$. Long dashes : HS mixture depletion potential ; Short dashes : truncated potential, $\phi^*$. Inset : reduced effective potential (in unit of $k_BT$) for the HS mixture (dashes) and the Asakura Oosawa one (solid line), for $\rho_s^* = 0.2$ and $\rho_s^* = 0.8$.

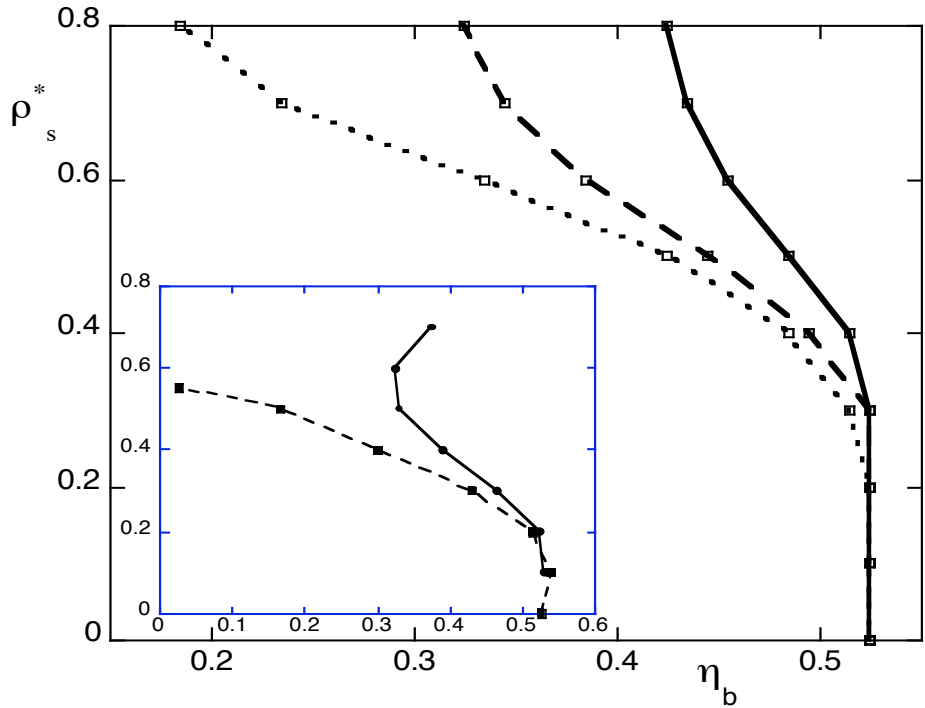

**Figure 3 :**

Gel transition lines of a mixture of « hard sphere-like » particles for q = 5, interacting through a residual Yukawa hetero-interaction with range $\xi_{sb} = \dfrac{\sigma_{sb}}{100}$. Dashes : pure hard spheres; Dots : $\varepsilon_{sb} = +\dfrac{3}{2}k_B T$ (with $\varepsilon_{sb}$ the contact value) ; $\varepsilon_{sb} = -\dfrac{3}{2}k_B T$. Inset : same for q = 10. Dashes : pure hard spheres ; Solid line : $\varepsilon_{sb} = -\dfrac{3}{2}k_B T$.